\pdfoutput=1

\documentclass[11pt,twoside,a4paper,cmspaper,final,collab]{cms-tdr}
\def\svnVersion{2297ffb-D}\def\svnDate{2019/09/20}\def\cmsCernNoTag{CERN-EP-2019-109}\def\cmsCernDate{\today}\def\cmsMessage{"Published in Physics Letters B as \href{http://dx.doi.org/10.1016/j.physletb.2019.134992}{\doi{10.1016/j.physletb.2019.134992}.}"}

\begin{document}\cmsNoteHeader{HIG-18-010}

\hyphenation{had-ron-i-za-tion}
\hyphenation{cal-or-i-me-ter}
\hyphenation{de-vices}
\RCS$ssHeadURL: svn+ssh://svn.cern.ch/reps/tdr2/papers/HIG-18-010/trunk/HIG-18-010.tex $
\RCS$Id$

\newlength\cmsFigWidth
\ifthenelse{\boolean{cms@external}}{\setlength\cmsFigWidth{0.49\textwidth}}{\setlength\cmsFigWidth{0.65\textwidth}} 
\ifthenelse{\boolean{cms@external}}{\providecommand{\cmsLeft}{upper\xspace}}{\providecommand{\cmsLeft}{left\xspace}}
\ifthenelse{\boolean{cms@external}}{\providecommand{\cmsRight}{lower\xspace}}{\providecommand{\cmsRight}{right\xspace}}
\newcommand{\PA}{\ensuremath{\cmsSymbolFace{A}}\xspace}
\newlength\cmsTabSkip\setlength{\cmsTabSkip}{1ex}
\providecommand{\CL}{CL\xspace}
\ifthenelse{\boolean{cms@external}}{\providecommand{\cmsTable}[1]{#1}}{\providecommand{\cmsTable}[1]{\resizebox{\textwidth}{!}{#1}}}

\cmsNoteHeader{HIG-18-010}
\title{Search for MSSM Higgs bosons decaying to $\Pgmp\Pgmm$ in proton-proton collisions at $\sqrt{s}=13\TeV$}

\date{\today}

\abstract{
A search is performed for neutral non-standard-model Higgs bosons decaying to two muons in the context of the minimal supersymmetric standard model (MSSM). Proton-proton collision data recorded by the CMS experiment at the CERN Large Hadron Collider at a center-of-mass energy of 13\TeV were used, corresponding to an integrated luminosity of 35.9\fbinv. The search is sensitive to neutral Higgs bosons produced via the gluon fusion process or in association with a $\cPqb\cPaqb$ quark pair. No significant deviations from the standard model expectation are observed. Upper limits at 95\% confidence level are set in the context of the $m_{\Ph}^{\text{mod+}}$ and phenomenological MSSM scenarios on the parameter $\tan\beta$ as a function of the mass of the pseudoscalar \PA boson, in the range from 130 to 600\GeV. The results are also used to set a model-independent limit on the product of the branching fraction for the decay into a muon pair and the cross section for the production of a scalar neutral boson, either via gluon fusion, or in association with \cPqb quarks, in the mass range from 130 to 1000\GeV.} 

\hypersetup{
pdfauthor={CMS Collaboration}, 
pdftitle={Search for MSSM Higgs bosons decaying to mu+mu- in proton-proton collisions at sqrt(s)=13 TeV},
pdfsubject={CMS},
pdfkeywords={CMS, Higgs, Muon, BSM, MSSM, model independent}}

\maketitle

\section{Introduction}
\label{sec:Introduction}
The boson discovered at the Large Hadron Collider (LHC) in 2012 \cite{bib:higg1,bib:higg2, bib:Higgs125}, with a mass around 125\GeV \cite{bib:combined1}, has properties that are consistent with those predicted for the standard model (SM) Higgs boson \cite{bib:combined2}. However, the SM is known to be incomplete, and several well-motivated theoretical models beyond the SM predict an extended Higgs sector. One example is supersymmetry \cite{Golfand:1971iw,Wess:1974tw} that protects the mass of the Higgs boson against quadratically divergent quantum corrections. In the minimal supersymmetric standard model (MSSM) \cite{bib:h_extension1,bib:h_extension2,bib:h_anatomy1}, the Higgs sector consists of two Higgs doublets, one of which couples to up-type fermions and the other to down-type fermions. Assuming that CP symmetry is conserved, this results in two charged bosons \Hpm, two neutral scalar bosons, \Ph and \PH, and one pseudoscalar boson, \PA.

At the tree level, the Higgs sector in the MSSM can be described by only two parameters, which are commonly chosen as $m_{\PA}$, the mass of the neutral \PA, and $\tan\beta$, the ratio of the vacuum expectation values of the neutral components of the two Higgs doublets. The masses of the other four Higgs bosons can be expressed as a function of these two parameters. Beyond the tree level the MSSM Higgs sector depends on additional parameters, which enter via higher-order corrections in perturbation theory, and which are usually fixed to values motivated by experimental constraints and theoretical assumptions. Setting these parameters defines a benchmark scenario \cite{bib:scenarios}, which is then described by $m_{\PA}$ and $\tan\beta$. The relevant scenarios are those consistent with a mass of one neutral boson of 125\GeV for the majority of the probed $m_{\PA}$--$\tan\beta$ parameter space \cite{bib:h_extension3}, and not ruled out by other existing measurements. In particular, the $m_{\Ph}^{\text{mod+}}$ scenario \cite{bib:scenarios} constrains the mass of the \Ph boson to be near 125\GeV for a wide range of $\tan\beta$ and $m_{\PA}$ values, by tuning some of the MSSM parameters. In the phenomenological MSSM (hMSSM) \cite{bib:hMSSM1,bib:hMSSM2,bib:hMSSM3,Djouadi:2013vqa} the mass of \Ph boson is an input parameter, set to 125\GeV, and the observed neutral boson is interpreted as the \Ph boson. Small differences in the cross sections and branching fractions exist between the two models, although the kinematics of the Higgs bosons remains almost identical.

This Letter reports on a search for beyond-the-SM neutral Higgs bosons in the dimuon final state in proton-proton ($\Pp\Pp$) collisions at a center-of-mass energy $\sqrt{s}$ of 13\TeV. The search is performed in the context of the MSSM for values of $m_{\PA}$ larger than 130\GeV, assuming either the $m_{\Ph}^{\text{mod+}}$ or the hMSSM scenario. For values of $m_{\PA} \gtrsim 200\GeV$, the MSSM is close to the decoupling limit: the \Ph boson takes the role of the observed SM-like Higgs boson at 125\GeV, and the \PH and \PA bosons are nearly degenerate in mass. For values of $m_{\PA}\lesssim 200\GeV$ the MSSM leads to similar, but not degenerate, masses for the \PH and \PA bosons \cite{bib:newScenarios}. The mass of the \Ph boson is assumed to be at 125\GeV, and its width smaller than the experimental resolution, consistently with the ATLAS and CMS measurements in other decay modes \cite{bib:combined1,bib:cmsWidth,bib:atlasWidth}. The analysis tests the \Ph boson production as predicted by the MSSM and the constraints on its production mechanisms measured by ATLAS and CMS are not enforced. Alternatively, the search is also performed in a model-independent way, where a neutral boson is assumed to be produced either via gluon fusion or in association with a $\cPqb\cPaqb$ quark pair. 

\begin{figure*}
\centering
\includegraphics[width=0.32\textwidth]{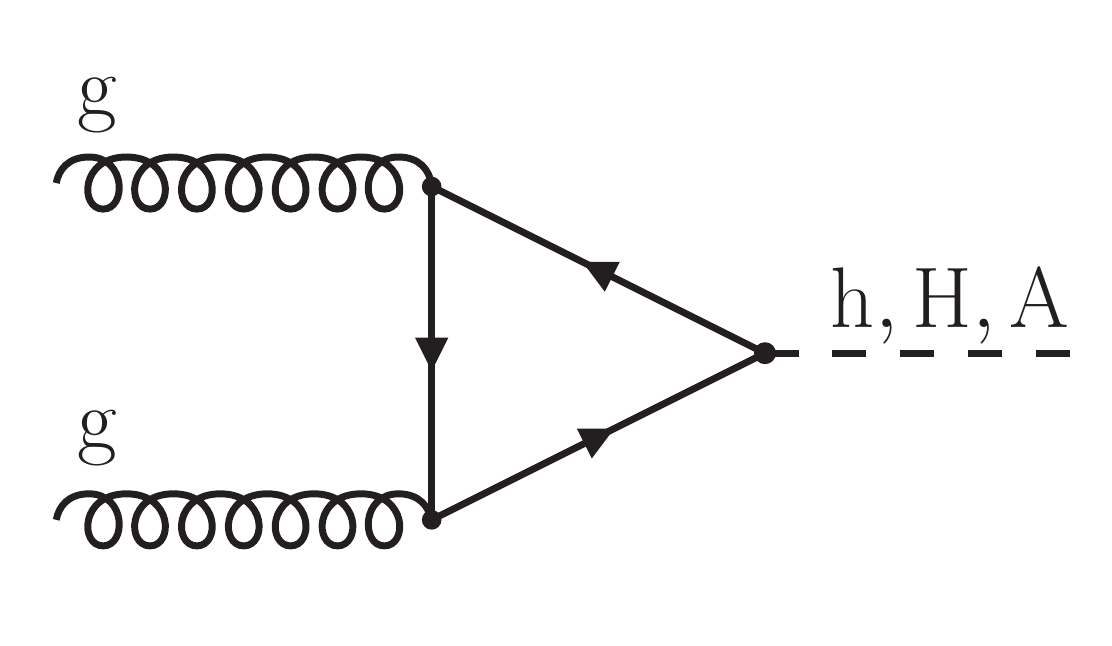}
\includegraphics[width=0.32\textwidth]{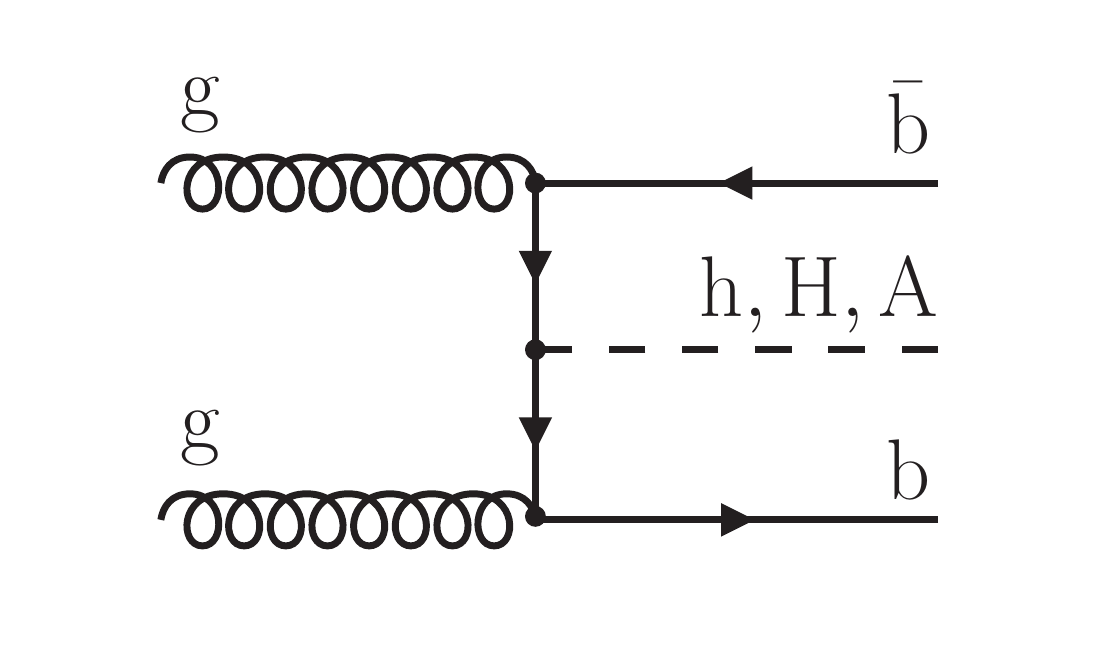}
\includegraphics[width=0.32\textwidth]{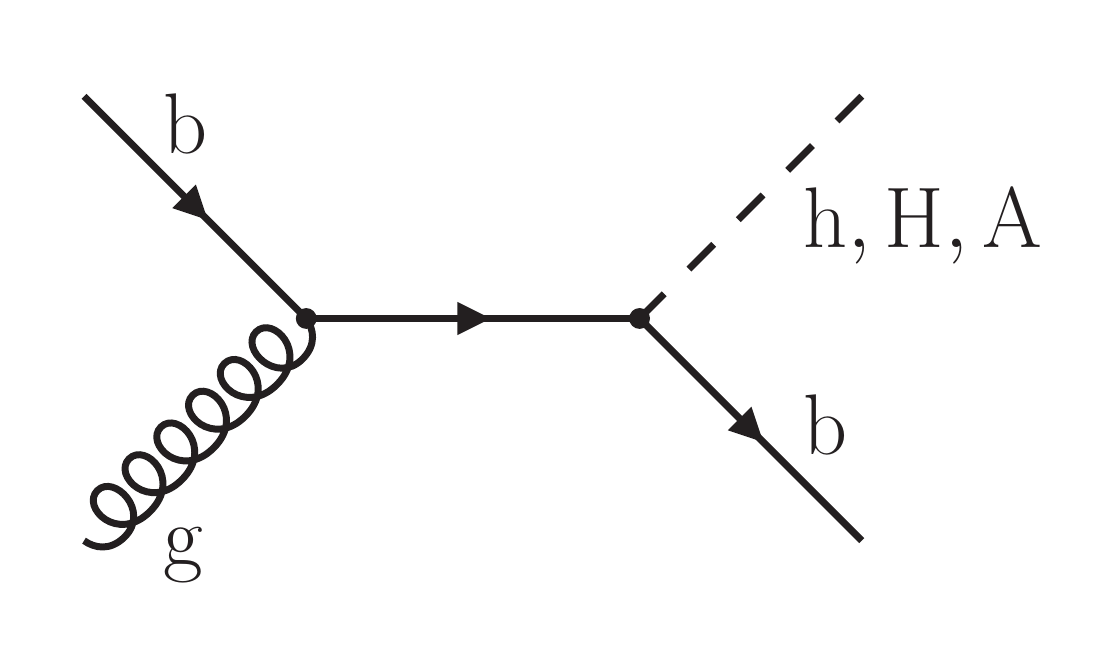}
\caption{Leading order Feynman diagrams for the production of the MSSM Higgs boson: gluon fusion production (left) and \cPqb-associated production (middle and right).}
\label{fig:Feynman}
\end{figure*}

At the LHC, dominant production mechanisms for the neutral \PA and \PH bosons are gluon fusion, in which the Higgs boson can be produced via a virtual loop of bottom or top quarks, and \cPqb-associated production, where the Higgs boson is produced in association with a \cPqb quark pair. This is also the case of the \Ph boson for values of $m_{\PA}\lesssim 200\GeV$, while, in the decoupling regime, the \Ph boson production mechanisms correspond to those predicted by the SM. Figure~\ref{fig:Feynman} shows the Feynman diagrams for the two production processes at leading order (LO). The gluon fusion mechanism is more relevant for $\tan\beta\lesssim 30$, whereas at LO, the coupling of the Higgs boson to down-type fermions is enhanced by $\tan\beta$, resulting in \cPqb-associated production becoming more important at large $\tan\beta$. The coupling of the neutral Higgs boson to charged leptons is enhanced for the same reason. Although the branching fraction to muons is predicted to be about 300 times smaller than that for the $\Pgt^{+}\Pgt^{-}$ final state, the $\Pgmp\Pgmm$ channel can be fully reconstructed, and the dimuon invariant mass can be measured with a precision of a few percent by exploiting the excellent muon momentum resolution of the CMS detector, making the dimuon final state an additional probe of the MSSM. 

The common experimental signature of the two production mechanisms is a pair of opposite-charge muons with high transverse momentum (\pt). The \cPqb-associated production process is characterized by the presence of additional jets originating from \cPqb quark fragmentation, whereas the events containing jets from light quarks or gluons are linked to the gluon fusion production mechanism. The presence of a signal would be characterized by an excess of events over the SM background in the dimuon invariant mass corresponding to the value of the Higgs boson masses.

The analysis is performed using the data at $\sqrt{s}=13\TeV$ collected during 2016 by the CMS experiment at the LHC corresponding to an integrated luminosity of 35.9\fbinv. Similar searches in the dimuon final state were performed by the ATLAS and CMS Collaborations using data collected in $\Pp\Pp$ collisions at 7 and 8\TeV \cite{Aad:2012cfr,bib:cms-mssm-2mu}, and by ATLAS at 13\TeV \cite{ref:atlas2019}. Searches for neutral Higgs bosons in the framework of the MSSM were performed by the ATLAS and CMS experiments also in the $\Pgt^{+}\Pgt^{-}$ \cite{Aad:2012cfr,Aad:2014vgg,Aaboud:2016cre,Aaboud:2017sjh,Chatrchyan:2011nx,Chatrchyan:2012vp,bib:cms-mssm-tautau} and $\cPqb\cPaqb$ \cite{Chatrchyan:2013qga,Khachatryan:2015tra,bib:cms-mssm-bb} final states. Limits on the existence of the MSSM Higgs bosons were determined also in $\Pep\Pem$ collisions at $\sqrt{s}=91$--209\GeV at the CERN LEP \cite{Schael:2006cr} and in proton-antiproton collisions at $\sqrt{s}=1.96\TeV$ at the Fermilab Tevatron \cite{bib:CDF_1,bib:CDF_2,bib:D0_1,bib:D0_2}.

\section{The CMS detector}
\label{sec:detector}

The central feature of the CMS apparatus is a superconducting solenoid of 6\unit{m} internal diameter, providing a field of 3.8\unit{T}. Within the field volume are a silicon pixel and strip tracker, a crystal electromagnetic calorimeter (ECAL), and a brass and scintillator hadron calorimeter (HCAL), each composed of a barrel and two endcap sections. Muons are measured in gas-ionization detectors embedded in the steel return yoke of the magnet. The first level (L1) of the CMS trigger system uses information from the calorimeters and muon detectors to select events of interest. The high-level trigger processor farm decreases the L1 accept rate from around 100\unit{kHz} to about 1\unit{kHz} before data storage. A more detailed description of the CMS detector, together with a description of the coordinate system and main kinematic variables used in the analysis, can be found in Ref. \cite{bib:cms_det}.

\section{Signal and background simulation}
\label{sec:data_mc_samples}

Samples of Monte Carlo (MC) simulated events are generated to model the Higgs bosons signal for the two leading production processes. This is done for a large number of $m_{\PA}$ and $\tan\beta$ combinations, where $m_{\PA}$ spans the range from 130 to 1000\GeV and $\tan\beta$ is varied from 5 to 60. Higgs boson events are generated with a mass within $\pm3\Gamma$ of the nominal Higgs boson mass, where $\Gamma$ is the intrinsic width. The values of $\Gamma$ strongly depend on $m_{\PA}$ and $\tan\beta$, being, for example, $\Gamma =$ 0.2~(2.7)\% of the nominal Higgs boson mass at $m_{\PA} = 150$ (550)\GeV and $\tan\beta = 10$ (40). The signal samples are generated with \PYTHIA 8.212 \cite{bib:pythia8} at LO. Additional signal samples are generated at next-to-LO (NLO) for some mass points to estimate higher-order corrections: gluon fusion samples are produced with \POWHEG 2.0 \cite{bib:powheg}, while \cPqb-associated production samples are produced with \mbox{\MGvATNLO} \cite{Alwall:2014hca} using the four-flavor scheme. 

{\tolerance=600
Simulated background processes are used to optimize the event selection but not to model the background shape and normalization, which are determined directly from data. The most relevant SM background processes considered are Drell--Yan (DY) production, and single and pair production of top quarks, which can produce $\Pgmp\Pgmm$ pairs with large invariant mass. Other background sources are the diboson production processes, $\PW^{\pm}\PW^{\mp}$, $\PW^{\pm}\cPZ$, and $\cPZ\cPZ$, whose \mbox{contributions} are each smaller than 1\% for dimuon invariant masses larger than 130\GeV, the Higgs boson search region. The background samples are generated at NLO using \mbox{\MGvATNLO} and \POWHEG. Spin correlations in multiboson processes generated using \mbox{\MGvATNLO} are simulated using \textsc{MadSpin} \cite{Artoisenet:2012st}. The NNPDF~3.0~\cite{bib:pdf} parton distribution functions (PDFs) are used for all samples. The parton shower and hadronization processes are modeled by \PYTHIA with the CUETP8M1 \cite{bib:CUETP8M1} underlying event tune. 
\par}

Detector response is based on a detailed description of the CMS detector and is simulated with the \GEANTfour package \cite{bib:geant}. Additional $\Pp\Pp$ interactions in the same or nearby bunch crossings (pileup) are simulated by \PYTHIA. During the data taking period, the CMS experiment was operating with, on average, 23 inelastic $\Pp\Pp$ collisions per bunch crossing. The distribution of the number of additional interactions per bunch crossing in the simulation is weighted to match that observed in the data. 

The values of the Higgs boson masses, widths, and the Yukawa couplings are calculated as a function of $m_{\PA}$ and $\tan\beta$ following the LHC Higgs Cross Section Working Group prescriptions \cite{LHCHXSWG4, bib:2013tqa}, using the \textsc{FeynHiggs} 2.12.0 \cite{Heinemeyer:1998yj,Heinemeyer:1998np,Degrassi:2002fi,Frank:2006yh,Hahn:2013ria} program for the $m_{\Ph}^{\text{mod+}}$ scenario. The inclusive cross sections of the Higgs bosons for the gluon fusion process are obtained with \textsc{SusHi} \cite{bib:sushi}, which includes NLO supersymmetric-QCD corrections \cite{Spira:1995rr,Harlander:2004tp,Harlander:2005rq,Degrassi:2010eu,Degrassi:2011vq,Degrassi:2012vt}, next-to-NLO (NNLO) QCD corrections for the top-quark contribution in the effective theory of a heavy top quark \cite{Harlander:2002wh,Anastasiou:2002yz, Ravindran:2003um,Harlander:2002vv, Anastasiou:2002wq}, and electroweak effects by light quarks \cite{Aglietti:2004nj, Bonciani:2010ms}. Higgs boson cross sections for the \cPqb-associated production are calculated with \textsc{SusHi}, and rely on matched predictions \cite{Harlander:2011aa}, which are based on the five flavour NNLO QCD calculation \cite{Harlander:2003ai} and the four flavour NLO QCD calculation \cite{Dittmaier:2003ej,Dawson:2003kb}. Higgs to $\Pgmp\Pgmm$ branching fractions are calculated with \textsc{FeynHiggs} for the $m_{\Ph}^{\text{mod+}}$ scenario and using the program \textsc{hdecay} 6.40 \cite{bib:hdecay} for the hMSSM scenario. Cross sections for the \ttbar and DY background processes are computed at the NNLO with \textsc{Top++2.0} \cite{TOPPP} and \FEWZ{3.1}~\cite{FEWZ}, respectively, while for the single top and the diboson production processes they are computed at NLO with \textsc{hathor} \cite{HATHOR2,HATHOR1} and \MCFM \cite{Campbell2011}, respectively.

\section{Object reconstruction and event selection}
\label{sec:object_reco}

The particle-flow (PF) algorithm \cite{bib:ParticleFlow} aims at reconstructing and identifying each individual particle in an event, with an optimized combination of information from the various elements of the CMS detector. The energy of photons is obtained from the ECAL measurement. The energy of electrons is obtained from a combination of the electron momentum at the primary interaction vertex as determined by the tracker, the energy of the corresponding ECAL cluster, and the energy sum of all bremsstrahlung photons spatially compatible with originating from the electron track. The energy of muons is obtained from the curvature of the corresponding track. The energy of charged hadrons is determined from a combination of their momentum measured in the tracker and the matching ECAL and HCAL energy deposits, corrected for zero-suppression effects and for the response function of the calorimeters to hadronic showers. Finally, the energy of neutral hadrons is obtained from the corresponding corrected ECAL and HCAL energies.

Muons with $20 < \pt < 100\GeV$ are measured with a relative \pt resolution of 1.3 to 2\% in the barrel and better than 6\% in the endcaps. The \pt resolution in the barrel is better than 10\% for muons with \pt up to 1\TeV \cite{bib:muonreco,bib:muonreco13TeV}.

Jets are reconstructed using the anti-\kt clustering algorithm \cite{Cacciari:2008gp} with a distance parameter of 0.4, as implemented in the \FASTJET package \cite{Cacciari:2011ma}. The quantity missing transverse momentum, \ptmiss, is defined as the magnitude of the negative vector \pt sum of all the PF objects (charged and neutral) in the event, and is modified by corrections to the energy scale of reconstructed jets. Collision vertices are obtained from reconstructed tracks using a deterministic annealing algorithm \cite{bib:vertexreco}. The reconstructed vertex with the largest value of summed physics-object $\pt^2$ is taken to be the primary $\Pp\Pp$ interaction vertex (PV). The physics objects are the jets, clustered using the jet finding algorithm~\cite{Cacciari:2008gp,Cacciari:2011ma} with the tracks assigned to the vertex as inputs, and the associated missing transverse momentum taken as the negative vector sum of the \pt of those jets.

The combined secondary vertex algorithm of Ref. \cite{bib:btag} is used to identify jets resulting from the hadronization of \cPqb quarks. A medium operating working point of the algorithm is applied to jets with $\pt > 20\GeV$ in the pseudorapidity range $\abs{\eta} < 2.4$. Within this kinematic range, the efficiency of the algorithm is 66\% with a misidentification probability of 1\%. 

The events are preselected by the trigger system \cite{bib:CMStrigger} requiring a muon candidate with $\abs{\eta} < 2.4$, satisfying at least one of the following criteria: $\pt > 24\GeV$ with isolation (iso) requirements, or $\pt > 50\GeV$ without isolation requirements. These are the trigger algorithms with the lowest \pt threshold whose output is not artificially reduced to limit the event rate and that cover the entire $\eta$ acceptance of the muon detector. Since the Higgs boson signal is searched for over a large mass range, the \pt of the muons from its decay can vary from tens to hundreds of GeV. Therefore, two sets of muon identification (ID) criteria are employed in the analysis: one is optimized for muons with lower \pt ($\lesssim 200\GeV$) (ID1) and the other for muons with larger \pt (ID2).

Events with a pair of opposite-charge muons, coming from the PV, are selected requiring both muons to satisfy the same ID criterion. Accepting, more generally, pairs of muons that pass any of the two ID criteria would lead to a negligible increase in signal efficiency. At least one of the two muon candidates has to match (in $\eta$ and azimuthal angle $\phi$ in radians) the muon that triggered the event. The trigger requirement depends on the ID algorithm. Offline reconstructed muons with $\abs{\eta} < 2.4$ are considered. Their offline \pt is required to be higher than 26 or 53\GeV, to be compatible with the muon that triggered the event. To reject muons from nonprompt decays, muon candidates must be isolated. The offline isolation variable is calculated depending on the ID algorithm, and is labelled iso1 (iso2) for ID1 (ID2). For ID1 it is the scalar \pt sum of the PF charged and neutral hadrons in a cone of radius $\Delta R = \sqrt{\smash[b]{(\Delta\eta)^2+(\Delta\phi)^2}} = 0.4$ around the muon direction, divided by the muon \pt. The charged PF particles not associated with the PV are not considered in this sum, and a correction is applied in order to account for the neutral particle contamination arising from pileup \cite{bib:pileupCacciari}. For ID2 the offline iso is computed as the scalar \pt sum of tracks in the silicon tracker, excluding the muon, in a cone of radius $\Delta R  = 0.3$ around the muon direction, and divided by the muon \pt. Tracks not associated with the PV are not considered. Energy deposits in the calorimeters are not included, since electromagnetic showers can develop from photons radiated by a high-\pt muon. The invariant mass of the Higgs boson candidate is reconstructed from the two highest-\pt opposite-charge muon candidates in the event. The dimuon selection criteria are summarized in Table~\ref{tab:muons}.

\begin{table*}[htb]
\centering
\topcaption{Summary of the muon selection criteria.}
\begin{tabular}{ c  c  c }
\hline 
Muon selection                        & muon ID1                        & muon ID2 \\
\hline
Online selection:                    & $\abs{\eta}< 2.4$               & $\abs{\eta}< 2.4$\\
Single muon                          & $\pt > 24\GeV$                  & $\pt > 50\GeV$ \\
                                     & Online iso                      & \\[\cmsTabSkip]
Offline selection:                        & $\abs{\eta}< 2.4$          & $\abs{\eta}< 2.4$ \\
Two opposite-charge muons                 & $\pt > 26\GeV$             & $\pt > 53 \GeV$\\ 
                                          & Offline iso1 $<$ 0.25      & offline iso2 $<$ 0.1\\ 
\hline
\end{tabular}
\label{tab:muons}
\end{table*}

The muon momentum measurement is crucial for the reconstruction of the Higgs boson mass peaks since improving the dimuon mass resolution increases the sensitivity of the analysis. To set limits accurately, the mean and the resolution of the dimuon mass peaks in simulation must match those of the data. A correction of the muon momentum has been applied in order to provide consistent measurements in the different $\phi$ and $\eta$ regions of the detector, improving the net resolution in data. The correction \cite{bib:muonreco13TeV} is also applied to the simulated muons to align the scale and resolution to those measured in the data. The magnitudes of the momentum scale corrections are about 0.2 and 0.3\% in the barrel and endcaps, respectively, for muons with \pt up to 200\GeV. For muons with larger \pt, since the statistical precision of the data is too poor to derive a correction, only a systematic uncertainty is considered (see Section \ref{sec:signal}).    

\begin{figure}[htb]
\centering
\includegraphics[width=0.49\textwidth]{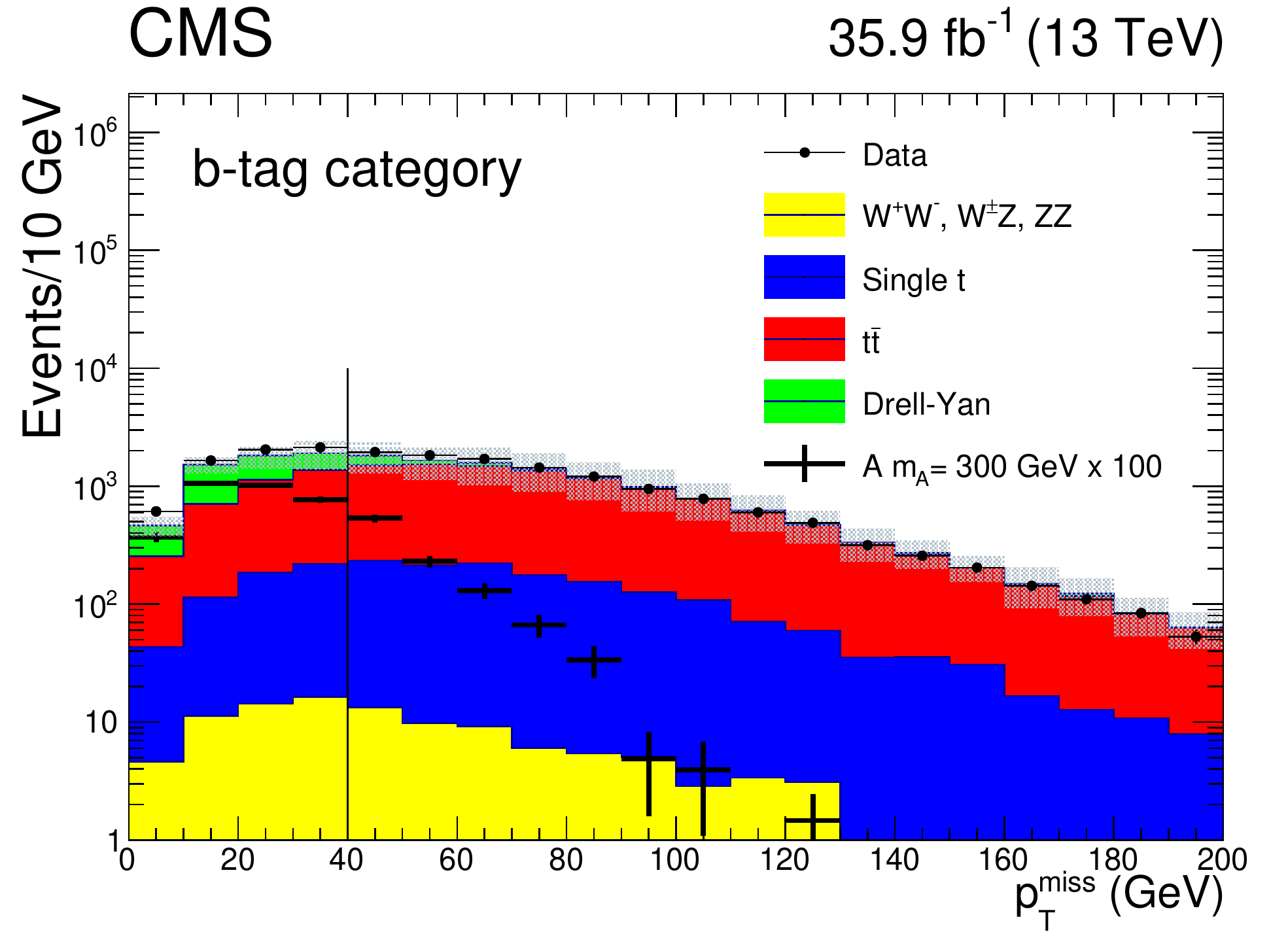}
\includegraphics[width=0.49\textwidth]{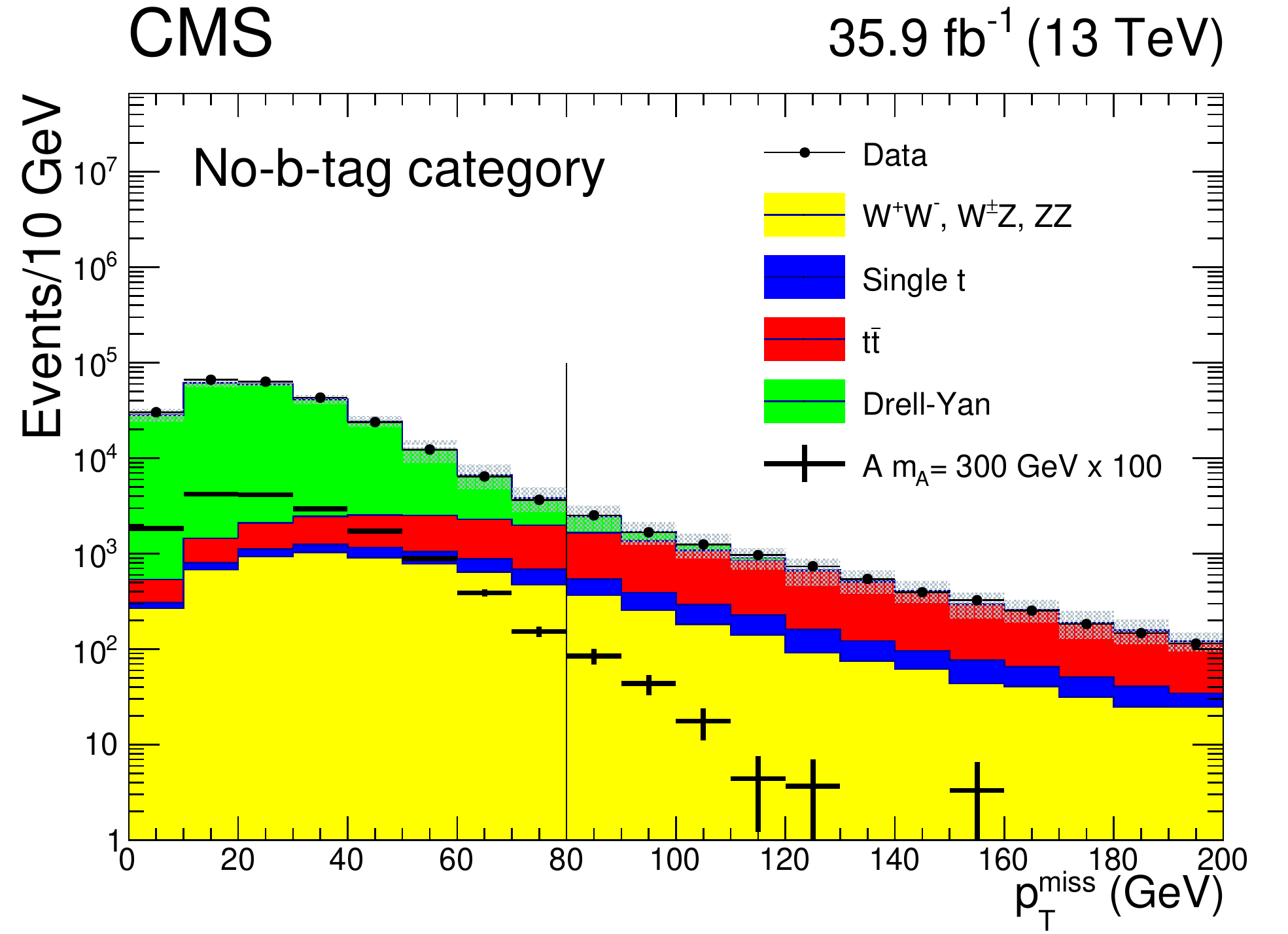}
\caption{Distribution of the missing transverse momentum in (\cmsLeft) \cPqb-tag and (\cmsRight) no-\cPqb-tag categories, for events with dimuon invariant mass larger than 130\GeV, as observed in data (dots) and predicted by simulation (colored histograms). The shaded gray band around the total background histogram represents the total uncertainty in the simulated prediction. The contribution of the expected signal for $m_{\PA} = 300\GeV$ and $\tan\beta = 20$, scaled by a factor of 100, is superimposed for illustration. The vertical line represents the upper threshold used to select the events in the two categories.}
\label{fig:control_met}
\end{figure}

When the Higgs boson is produced in association with a $\cPqb\cPaqb$ pair, additional jets from $\cPqb$ quark fragmentation are expected. Jets with $\pt > 20\GeV$ and $\abs{\eta}< 2.4$ are considered in this analysis: those that satisfy the requirements for the medium \cPqb-tagging working point \cite{bib:btag} are taken as \cPqb-jet candidates, otherwise they are taken as untagged jets. Events containing \cPqb-jet candidates provide the highest sensitivity for the \cPqb-associated production channel, and events that do not contain \cPqb-tagged jets provide the best sensitivity for the gluon fusion production channel. The events are therefore split into two exclusive categories: the \cPqb-tag category, containing events with strictly one \cPqb jet and at most one additional untagged jet, and the no-\cPqb-tag category, containing events without \cPqb-tagged jets. In the first category, the requirement of strictly one \cPqb jet is aimed at suppressing about 30\% of the dominant background from top quark pairs, since the observed \cPqb-tagged jet multiplicity in $\cPqt\cPaqt$ events is on average higher than for the Higgs boson signal. This is because more than half of the signal events from \cPqb-associated production are characterized by \cPqb jets emitted at large $\eta$, out of the acceptance of the tracking detector, and failing the \cPqb-tag requirements, whereas \cPqb jets in $\cPqt\cPaqt$ events are preferentially emitted in the central $\eta$ region. Therefore, discarding events with two or more \cPqb-tagged jets allows the $\cPqt\cPaqt$ background to be rejected without any major impact on the signal efficiency. Furthermore, $\cPqt\cPaqt$ events are characterized by a higher multiplicity of additional untagged jets than the signal events.

Signal events are characterized by a rather small \ptmiss. However, the background content is quite different for the two categories, as shown in Fig.~\ref{fig:control_met}. The background from $\cPqt\cPaqt$ events, characterized by a relatively large \ptmiss from \PW boson decays, is much more relevant for the \cPqb-tag category. For the no-\cPqb-tag category, the dominant background is DY production, whose events are characterized by a \ptmiss distribution that is similar to that of the signal. For this reason, a requirement on \ptmiss, separately tuned for the \cPqb-tag and the no-\cPqb-tag events, improves the background rejection and increases the signal sensitivity. Events belonging to the \cPqb-tag (no-\cPqb-tag) category are required to have $\ptmiss<40$ (80)\GeV. This requirement reduces the background from top quark production by about 75\% (40\%). The selection criteria that define the two categories are summarized in Table~\ref{tab:categories}.

\begin{table*}
\caption{Summary of the selection criteria that define the two event categories. Categorization is applied after the muon selection.}
\centering
\begin{tabular}{ c  c  c }
\hline
                            & \cPqb-tag category            & No-\cPqb-tag category \\
\hline
\cPqb-tagged jets               & 1 with $\pt > 20\GeV$, $\abs{\eta} < 2.4$   & Veto \\
Untagged jets               & 0,1 with $\pt > 20\GeV$, $\abs{\eta} < 2.4$ & \\
\ptmiss                     & $<$40\GeV                                   & $<$80\GeV \\ 
\hline
\end{tabular}
\label{tab:categories}
\end{table*}

\section{Signal efficiency and signal systematic uncertainties}
\label{sec:signal}    

For each value of $m_{\PA}$ and $\tan\beta$, the signal efficiency for each Higgs boson sample is defined as the fraction of generated events that fulfill the selection criteria. This definition of efficiency also includes the effects of limited detector acceptance and the selections outlined in Section \ref{sec:object_reco}. 

\begin{figure}
\centering
\includegraphics[width=0.49\textwidth]{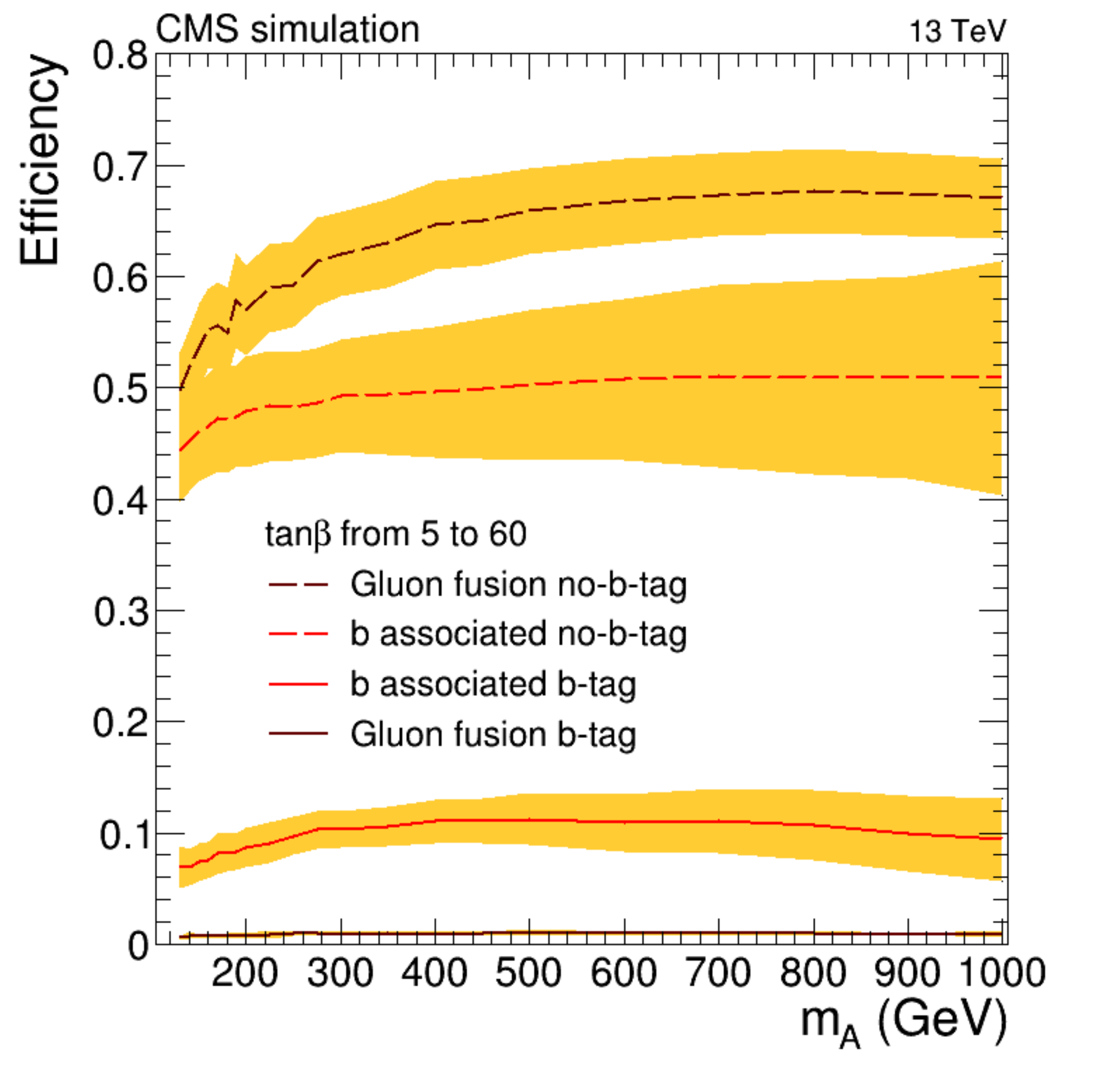}
\caption{The selection efficiency for the \PA boson, as a function of its mass, for the two production mechanisms, \cPqb-associated and gluon fusion, and for each of the two event categories. The band centered on each curve corresponds to the envelope of efficiencies obtained when varying $\tan\beta$, combined with the statistical and systematic uncertainties. }
\label{fig:Aeffic}
\end{figure}

Figure~\ref{fig:Aeffic} shows the selection efficiency for the \PA boson as a function of $m_{\PA}$, for the gluon fusion and the \cPqb-associated production processes, and for the two event categories. Each curve corresponds to the mean of the efficiency obtained by varying  $\tan\beta$ between 5 and 60, while the band of each curve corresponds to the efficiency variations combined with the statistical and systematic uncertainties (described in the next paragraph) of the simulated samples. For a given mass, the selection efficiency is weakly dependent on $\tan\beta$, since this parameter mostly affects the Higgs boson width, with a negligible impact on the kinematic properties of the event. The efficiency to detect events produced in association with \cPqb quarks is approximately 10\% at high masses for the \cPqb-tag category. This value is mostly determined by the large fraction of \cPqb jets that are emitted with an $\eta$ value that is outside the coverage of the tracking detectors, and indeed $\approx$50\% of events from \cPqb-associated samples are reconstructed in the no-\cPqb-tag category. The efficiency to detect events from gluon fusion reaches a maximal value at $\approx$65\% for $m_{\PA}\gtrsim 400\GeV$. The very small but nonvanishing efficiency for signal produced via gluon fusion in the \cPqb-tag category is due to the \cPqb misidentification probability, which is about 1\%. The corresponding efficiencies for the \PH boson are consistent with those shown in Fig.~\ref{fig:Aeffic}.

\begin{table*}[htb]
\topcaption{Systematic uncertainties in the signal efficiency for the two event categories. The systematic uncertainties hold for both Higgs boson production processes except for the sources listed in the last three rows, which apply to the \cPqb-associated production process only. For these three sources, in the model-independent search for a neutral boson produced in association with \cPqb quarks, the uncertainties are applied as quoted in the table. In the MSSM interpretation, these numbers have to be weighted by the relative contribution of the \cPqb-associated production process to each category. For those sources of systematics that depend on $m_{\PA}$ the range of uncertainty is quoted.}
\centering
\cmsTable
{
\begin{tabular}{ l  c  c }
\hline
Source                                    & \multicolumn{2} {c} {Systematic uncertainty (\%)} \\ 
                                          & \cPqb-tag category     & No-\cPqb-tag category    \\
\hline
MC statistical uncertainty                & 0.5--6    & 0.2--2                              \\ 
Trigger efficiency                        & 0.9       & 0.9                                 \\
Muon reconstruction                       & 2         & 2                                   \\ 
Muon isolation                            & 1         & 2                                   \\ 
Pileup                                    & 0.8       & 0.9                                 \\ 
Jet energy scale                          & 1.6       & 0.4                                 \\ 
Unclustered energy                        & 4.1       & 0.3                                 \\ 
PDF                                       & 3         & 3                                   \\ 
Higgs boson \pt                           & 1--4      & 1--4                                \\[\cmsTabSkip]
\cPqb tag (only for \cPqb-associated production)  & 2         & 0.6                                 \\
\cPqb jet multiplicity (only for \cPqb-associated production) & 20--30 & 7--20                      \\ 
Untagged jet multiplicity (only for \cPqb-associated production) & 7--25 &   \NA                \\ 
\hline
\end{tabular}
}
\label{tab:sys}
\end{table*}

The systematic uncertainties in the signal description arise from a possible mismodeling of the signal efficiency, of the signal shape, and, for the model interpretation, from uncertainties in its cross section.

The systematic uncertainties that affect the signal efficiency are given in Table~\ref{tab:sys}. The size of the simulated signal samples introduces a statistical uncertainty in the signal efficiency that is between 0.2\% and 6\%, depending on the number of generated events. 

In order to account for the differences between data and simulation in the muon trigger efficiency, identification, and isolation, scale factors calculated using the tag-and-probe technique \cite{bib:muonreco,bib:muonreco13TeV} have been applied to simulated events. A similar procedure is used to account for discrepancies between data and simulation in the \cPqb-tagging efficiency. A global correction, calculated as the product of the various scale factors, is applied as an event-by-event weight. The uncertainty associated with each scale factor is then propagated to the analysis and its impact on the final selection efficiency is assigned as systematic uncertainty. An event-by-event weight is also applied to account for the modeling of the pileup in the simulation. The uncertainty in the knowledge of the pileup multiplicity is evaluated by varying the total inelastic cross section \cite{bib:CMS_ppxs,bib:ATLAS_ppxs} by $\pm5\%$, which translates into an uncertainty smaller than 1\% in the signal efficiency. The uncertainty associated with the jet energy scale \cite{Chatrchyan:2011ds} is estimated by rescaling the jet momentum by a factor depending on the \pt and $\eta$ of each jet. This variation is also propagated to the \ptmiss determination. Its effect on the signal selection efficiency is about 1.6 (0.4)\% for the \cPqb-tag (no-\cPqb-tag) category. Systematic uncertainties in the unclustered energy are propagated to the determination of \ptmiss. The effect on the signal efficiency is 4.1\% for the \cPqb-tag category, and 0.3\% for the no-\cPqb-tag category. Systematic uncertainty in the \cPqb-tagging algorithm affects the signal yield and the category migration with an impact on the signal efficiency of 2\% for the \cPqb-tag category and 0.6\% for the no-\cPqb-tag category. The uncertainty in the total integrated luminosity is 2.5\% \cite{bib:lumi} and affects the signal yield. 

The uncertainties in the MSSM cross sections depend on $m_{\PA}$, $\tan\beta$, and the scenario. They are provided by the LHC Higgs Cross Section Working Group \cite{LHCHXSWG4, bib:2013tqa}. An uncertainty of 3\% is used to account for the parton distribution functions. 

Additional corrections are applied to take into account the fact that the signal samples are generated with \PYTHIA at LO instead of using an NLO generator. Higher-order corrections affect the Higgs boson \pt modeling, with impacts on the muon acceptance and the jet multiplicity. Moreover, they cause event migration between the two categories. The acceptance obtained from the LO samples is corrected to that predicted at NLO. The corresponding systematic uncertainty is set to the size of the correction itself. The correction on the modeling of the Higgs \pt increases the signal efficiency by 1--4\%, depending on the Higgs boson mass. The correction on the \cPqb-jet multiplicity affects only the \cPqb-associated signal, resulting in a correction of 20--30\% depending on $m_{\PA}$, which increases the signal efficiency for the \cPqb-tag category, and a correction of 7--20\% decreasing the signal efficiency for the no-\cPqb-tag category. An additional correction of 7--25\%, related to the untagged jet multiplicity, is applied, and reduces the signal efficiency for the \cPqb-tag category, due to the veto on the untagged jets.    

The systematic uncertainties in the \cPqb-tag efficiency and the jet multiplicity shown in Table~\ref{tab:sys} apply only to the \cPqb-associated production process. Both the \cPqb-tagging and the \cPqb-jet multiplicity uncertainties are anticorrelated between the two event categories. In the model-independent analysis for the case in which the neutral boson is assumed to be entirely produced in association with \cPqb quarks, these uncertainties are applied, as quoted in Table~\ref{tab:sys}, while in the MSSM interpretation, where both the gluon fusion and the \cPqb-associated production processes contribute to the two event categories, these systematic uncertainties are weighted by the relative contribution of the latter process.

The shape of the reconstructed Higgs boson invariant mass distribution is affected by the muon momentum scale and resolution. Uncertainties in the calibration of these quantities are propagated to the shape of the invariant mass distribution assuming a Gaussian prior, leading to a variation of up to 10\% in the width of the signal mass peak, and to a negligible shift of its position. These uncertainties are taken into account as a signal shape variation in the calculation of the exclusion limit.

\section{Modeling of the signal and background shapes}
\label{sec:signalfit}
The invariant mass spectrum of the signal events that pass the event selection is used to determine the signal yield for each category. In the framework of the MSSM, this is done by fitting the invariant mass distribution of the \Ph, \PH, and \PA bosons, separately for the two event categories and for various combinations of $m_{\PA}$--$\tan\beta$ values. The function $F_{\text{sig}}$ used to parametrize the signal mass shape \cite{bib:cms-mssm-2mu} is defined as:
\begin{linenomath}
\begin{equation}
F_{\text{sig}} =  w_{\Ph}F_{\Ph} + w_{\PH}F_{\PH} +  w_{\PA}F_{\PA}.
\label{eq:signalfit}
\end{equation}
\end{linenomath}
In Eq.~(\ref{eq:signalfit}), the terms $F_{\Ph}$, $F_{\PH}$, and $F_{\PA}$ describe the mass shape of the \Ph, \PH, and \PA signals, respectively. Each term is a convolution of a Breit--Wigner (BW) function to describe the resonance, with a Gaussian function to account for the detector resolution. The two parameters of the BW function, as well as the variance of each Gaussian function, are free parameters of the fit used to determine the signal model, while the quantities $w_{\Ph}$, $w_{\PH}$, and $w_{\PA}$ are the numbers of expected events for each boson passing the event selection. For the $m_{\PA}$--$\tan\beta$ points for which the signal samples were not generated, the parameters are interpolated from the nearby generated points. In order to correct for differences of the order of a few\GeV between the \PYTHIA prediction of $m_{\PH}$ with respect to the value calculated by \textsc{FeynHiggs} in the $m_{\Ph}^{\text{mod+}}$ or the value used in the hMSSM, especially for $m_{\PA} \lesssim 200\GeV$, the invariant mass distribution of the \PH boson is shifted by the corresponding amount. For the model-independent analysis the signal shape is described using one single resonance in Eq.~(\ref{eq:signalfit}).

The analysis does not use background estimation from simulation due to the limited size of simulated events compared to data in the region of interest, as well as due to the large theoretical uncertainties in the background description at high invariant masses. Therefore, given the smooth dependence of the background shape on the dimuon invariant mass, it is estimated from the data, by assuming a functional form to describe its dependence as a function of the reconstructed dimuon invariant mass, $m_{\mu\mu}$, and by fitting it to the observed distribution.

The functional form used to describe the background shape is defined as:
\ifthenelse{\boolean{cms@external}}{
\begin{linenomath}
\begin{multline}
F_{\text{bkg}} =  \exp(\lambda m_{\mu\mu})\\
\times \left[\frac{f}{N_1}\frac{1}{{(m_{\mu\mu}-m_{\cPZ})}^2 +\frac{\Gamma^2_{\cPZ}}{4}} + \frac{(1-f)}{N_2}\frac{1}{{m_{\mu\mu}}^2} \right].
\label{eq:BWZgamma}
\end{multline}
\end{linenomath}
}{
\begin{linenomath}
\begin{equation}
F_{\text{bkg}} =  \exp(\lambda m_{\mu\mu}) \left[\frac{f}{N_1}\frac{1}{{(m_{\mu\mu}-m_{\cPZ})}^2 +\frac{\Gamma^2_{\cPZ}}{4}} + \frac{(1-f)}{N_2}\frac{1}{{m_{\mu\mu}}^2} \right].
\label{eq:BWZgamma}
\end{equation}
\end{linenomath}
}
The quantity $\exp(\lambda m_{\mu\mu})$ parametrizes the exponential part of the mass distribution, and $f$ represents the weight of the BW term with respect to DY photon exchange, while N$_{1}$ and N$_{2}$ correspond to the integral of each term in $F_{\text{bkg}}$. The quantities $\lambda$ and $f$ are free parameters of the fit. The parameters $\Gamma_{\cPZ}$ and $m_{\cPZ}$ are separately determined for the two event categories by fitting the dimuon mass distribution close to the $\cPZ$ boson mass. The fit provides the effective values of such quantities, which include detector and resolution effects. Their values are then kept constant when using $F_{\text{bkg}}$ in the final fit. The systematic uncertainty that stems from the choice of the functional form in Eq.~(\ref{eq:BWZgamma}), which was used in earlier searches \cite{bib:cms-mssm-2mu}, is assessed as described below.

A linear combination of the functions describing the expected signal and the background is then used to perform a binned maximum likelihood fit to the data, where the uncertainties are treated as nuisance parameters: 
\begin{linenomath}
\begin{equation}
F_{\text{fit}} = (1 - f_{\text{bkg}})F_{\text{sig}} + f_{\text{bkg}}F_{\text{bkg}}.
\label{eq:Ffit}
\end{equation}
\end{linenomath}
The fit is performed for each $m_{\PA}$ and $\tan\beta$ hypothesis, as the yield of the signal events and the shape of $F_{\text{sig}}$ depend on these quantities. The parameters that describe the signal are determined by fitting the simulated samples that pass the event selection with Eq.~(\ref{eq:signalfit}), for each $m_{\PA}$ and $\tan\beta$ pair, as explained above. Subsequently they are assigned as constant terms in $F_{\text{fit}}$. The quantity $f_{\text{bkg}}$ is a free parameter in the fit, and the fraction of signal events is defined as $f_{\text{sig}}=(1-f_{\text{bkg}})$. The overall normalization is also a free parameter and is profiled in the fit.

For each $m_{\PA}$ assumption, the function $F_{\text{fit}}$ is used to fit the data over an $m_{\mu\mu}$ range centered on $m_{\PA}$. The range has to be large enough to account for the signal width, including the experimental resolution, and it is $\pm 50\GeV$ for $m_{\PA} \leq 290\GeV$, $\pm75\GeV$ for $290 < m_{\PA} \leq 390\GeV$, and $\pm 100\GeV$ for $390 < m_{\PA}\leq 500\GeV$. For values of $m_{\PA}$ smaller than 165\GeV the lower bound of the mass window is set to 115\GeV. For $m_{\PA} > 500\GeV$, the entire range from 400 to 1200\GeV is used. The \Ph boson is used to constrain the results when its mass is included in the fitted mass range. 

The uncertainty introduced by the choice of the analytical function used to parametrize the background is estimated by using a method similar to that used in Refs. \cite{bib:Higgs125,bib:cms-mssm-2mu,Sirunyan:2018hbu}. The method is based on the determination of the number of spurious signal events that are introduced by the choice of the background function $F_{\mathrm{bkg}}$, when the background is fit by the function $F_{\mathrm{fit}}$. The invariant mass spectrum is fitted by the function $F_{\mathrm{bkg}}^a$, chosen among various functional forms: Eq.~(\ref{eq:BWZgamma}) or other similar expressions that include a BW plus exponentials, and sum of exponentials. All these functional forms adequately describe the background distribution observed in data. The fit is performed in the proper mass range centered around the assumed value of $m_{\PA}$, and the parameters of $F_{\mathrm{bkg}}^a$ are determined. Then, thousands of MC pseudo-experiments are generated, each one containing the same number of events as observed in the data, distributed according to the functional form $F_{\mathrm{bkg}}^a$.  For each pseudo-experiment, the invariant mass distribution is then fit with the function $F_{\mathrm{fit}}$ of Eq.~(\ref{eq:Ffit}), once using $F_{\mathrm{bkg}}^a$, and then using a different function $F_{\mathrm{bkg}}^b$, given by Eq.~(\ref{eq:BWZgamma}). For each pseudo-experiment, the spurious signal yield, expressed by the number of events $N_{\mathrm{bias}}^a$ and $N_{\mathrm{bias}}^b$, is determined. The quantity $N_{\mathrm{bias}}^a$ is on average consistent with zero within statistical fluctuations. The quantity $N_{\mathrm{bias}}^b$ represents the number of spurious signal events that are found in the signal yield if the function $F_{\mathrm{bkg}}^b$ is used to describe the background, when the background itself is actually distributed according to $F_{\mathrm{bkg}}^a$. The median of the distribution of the difference $N_{\mathrm{bias}}^a - N_{\mathrm{bias}}^b$ obtained from the pseudo-experiments is defined as the bias introduced by using the function $F_{\mathrm{bkg}}^b$, relative to the tested mass $m_{\PA}$. This procedure is repeated for each function $F_{\mathrm{bkg}}^a$ among the functional forms mentioned above, and the largest bias is taken as the systematic uncertainty in the number of signal events obtained from the maximum likelihood fit, due to the choice of Eq.~(\ref{eq:BWZgamma}) to parametrize the background distribution. Choosing a different function $F_{\mathrm{bkg}}^b$, instead of Eq.~(\ref{eq:BWZgamma}), was shown to lead to similar biases over the whole mass range. The number of spurious signal events varies between a few units and a few hundred depending on the mass of the signal and the event category. Although the bias is due to the modelling of the background, its impact on the result depends on the expected signal strength and shape, both varying according to $m_{\PA}$ and $\tan\beta$ in the model-dependent analysis, and according to the mass of a generic resonance $\phi$ for the model-independent case. More details about the effect of the bias on the final results are discussed in Section \ref{sec:Results}. 

\begin{figure*}
\centering
\includegraphics[width=0.49\textwidth]{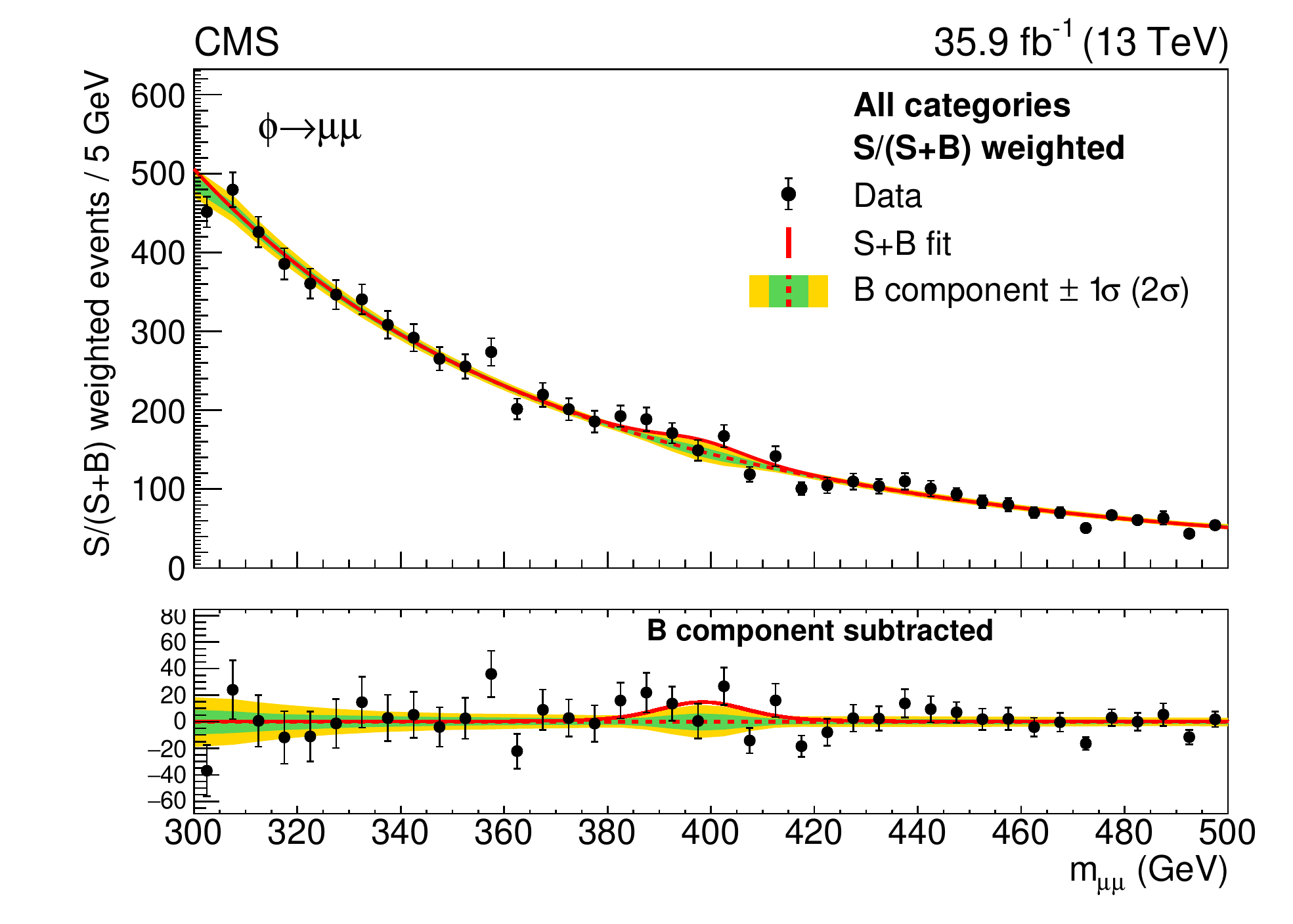}
\includegraphics[width=0.49\textwidth]{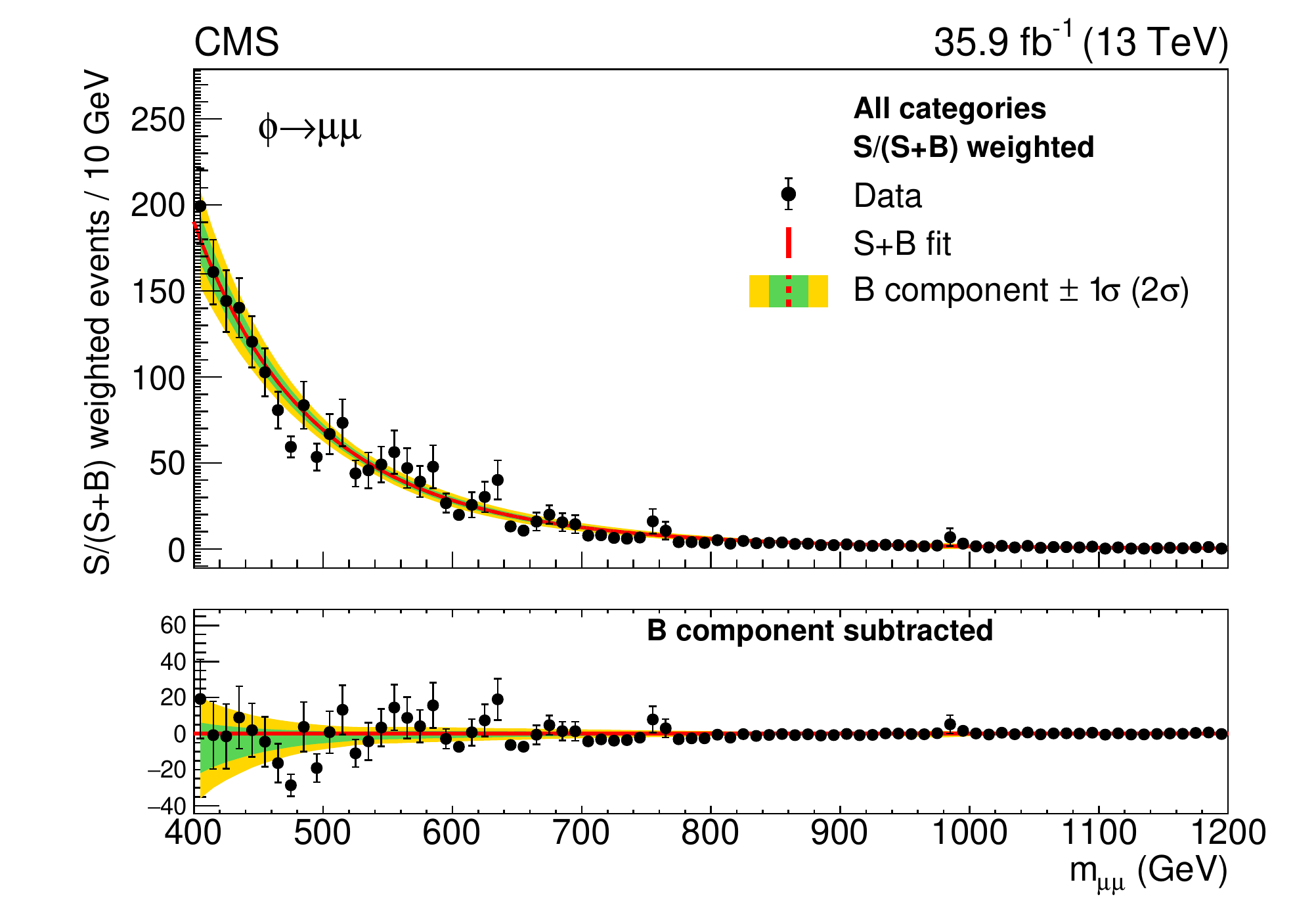}
\caption{Examples of fits to data with a signal plus background hypothesis, for a narrow-width signal with a mass of 400\GeV (left), and 980\GeV (right), for the two event categories added together, after weighting by their sensitivity. The resonance $\phi$ is assumed to be produced via the \cPqb-associated production, and to decay to two muons. The 68 and 95\% \CL bands, shown in dark green and light yellow, respectively, include the uncertainties in the background component of the fit. The lower panel shows the difference between the data and the background component of the fit.}
\label{fig:postfit}
\end{figure*}

An example of fits to the data with Eq.~(\ref{eq:Ffit}), for the model-independent case, is shown in Fig.~\ref{fig:postfit}. Two mass hypotheses, 400 and 980\GeV, are assumed for a single narrow-width resonance $\phi$ decaying to two muons. The two event categories are combined according to their sensitivity, $S/(S+B)$, where S and B are the number of events in the expected signal and observed background, respectively. The uncertainties in the integrated luminosity, in the signal efficiency, and in the background parametrization are taken into account as nuisance parameters. 

\section{Results}
\label{sec:Results}

No evidence of Higgs boson production beyond the SM prediction is observed in the mass range in which the analysis has been performed. Exclusion limits at 95\% confidence level (\CL) are therefore determined. 

A maximum likelihood fit to the data, as explained in the previous section, is performed under the background only and the signal-plus-background hypotheses, where the background includes the expectation for the SM Higgs boson. The systematic uncertainties are incorporated as nuisance parameters in the likelihood. The upper limits for the signal production are computed using the \CLs~\cite{bib:hybrid1,bib:hybrid2} criterion and the hybrid frequentist-bayesian approach, where the distributions of the test-statistic are derived from pseudo-experiments \cite{ATLAS:2011tau}.

The results are interpreted within the MSSM in the context of the $m_{\Ph}^{\text{mod+}}$ and hMSSM scenarios, by combining both event categories. The 95\% \CL limit on the parameter $\tan\beta$ is presented as a function of $m_{\PA}$: the exclusion limit is chosen for each $m_{\PA}$ as the $\tan\beta$ value at which the \CLs is lower than 0.05. 

\begin{figure}[htbp]
\centering
\includegraphics[width=0.49\textwidth]{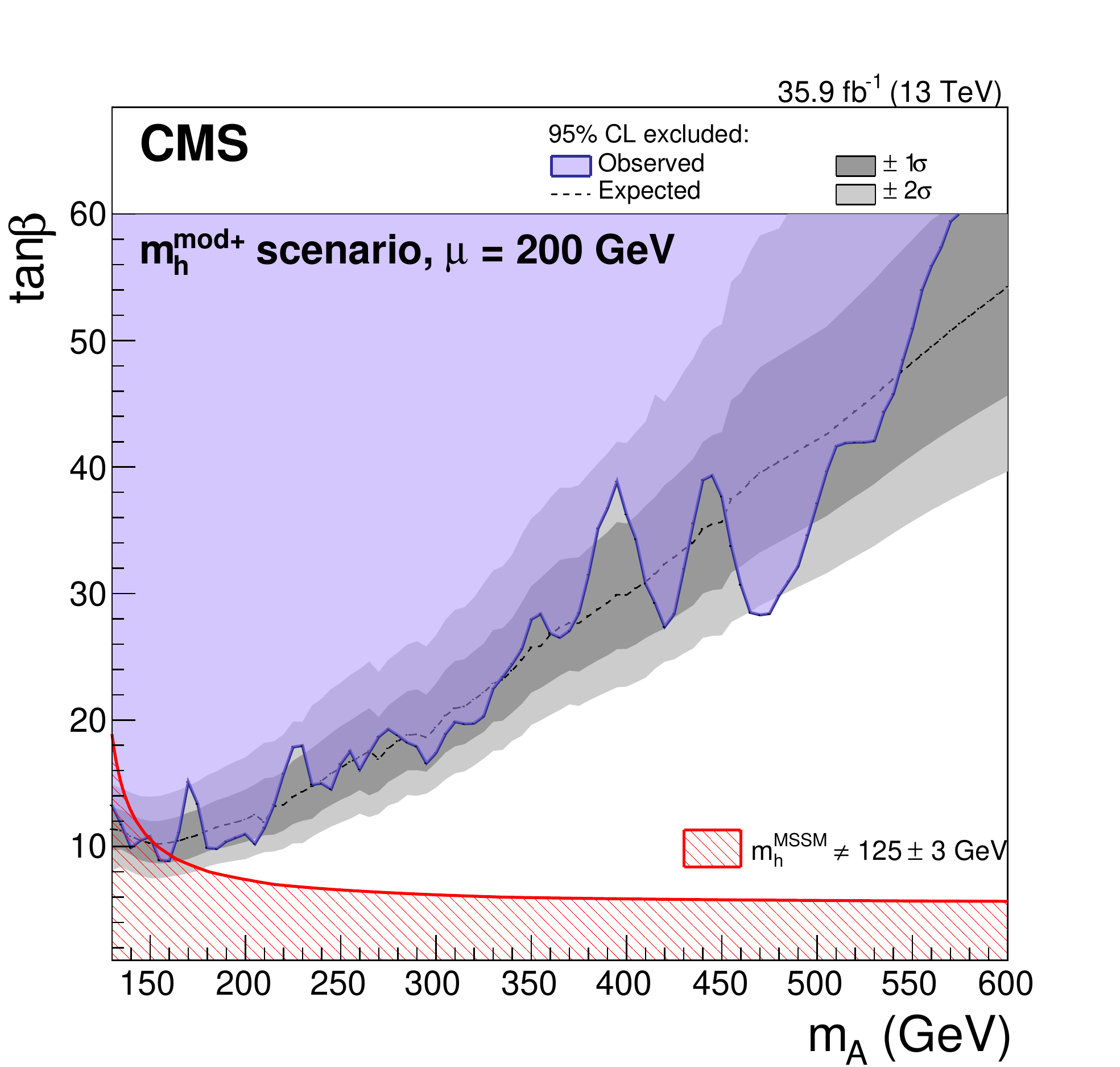}
\includegraphics[width=0.49\textwidth]{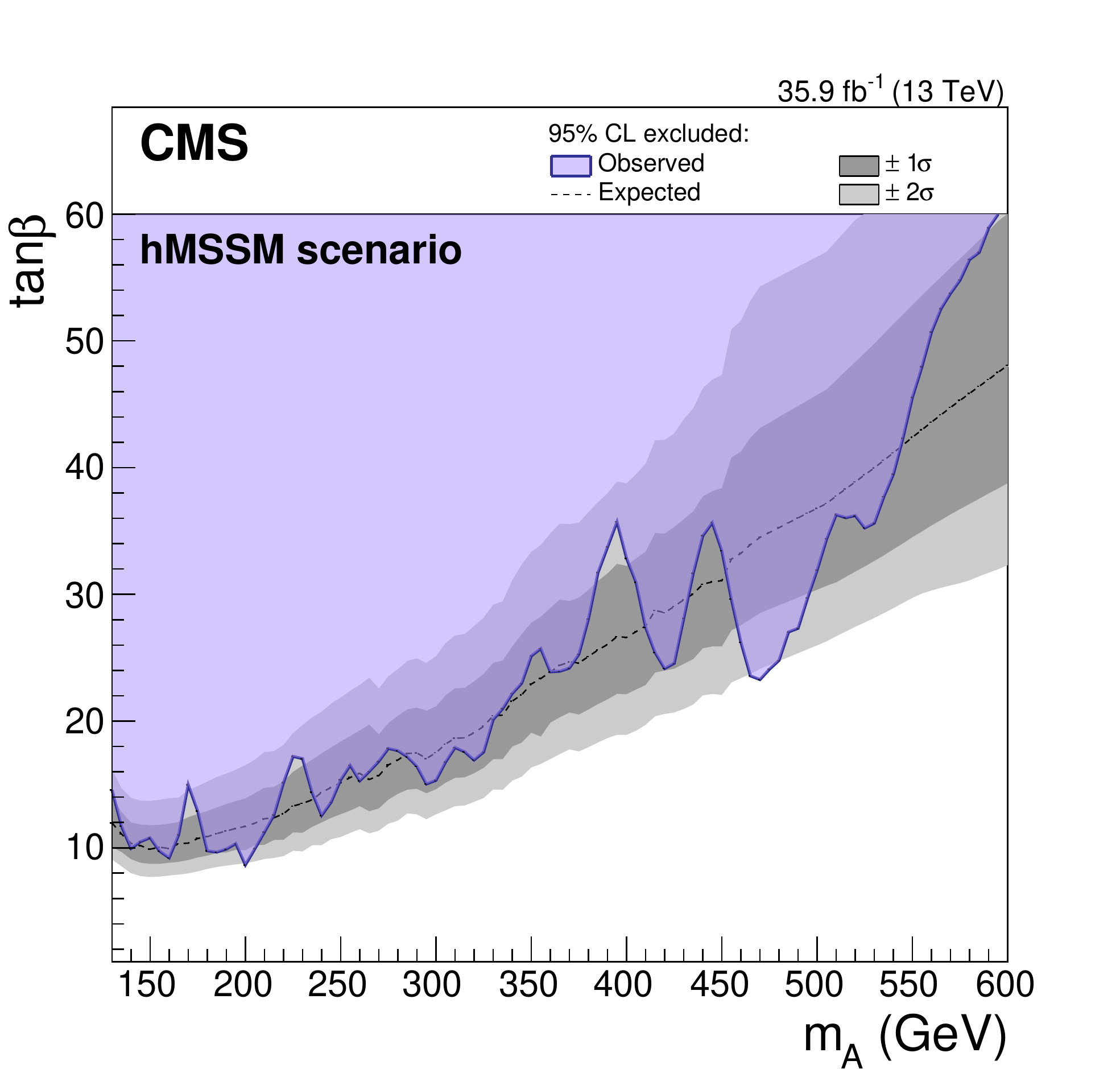}
\caption{The 95\% \CL expected, including the 68 and 95\% \CL bands, and observed upper limits, on $\tan\beta$ as a function of $m_{\PA}$ for the $m_{\Ph}^{\text{mod+}}$ (\cmsLeft) and the hMSSM (\cmsRight) scenarios of the MSSM. The observed exclusion contour is indicated by the purple region, while the area under the red curve is excluded by requiring the neutral \Ph boson mass consistent with $125\pm3\GeV$.}
\label{fig:MSSM_hMSSM_limits}
\end{figure}

To estimate the impact of the various systematic uncertainties, the 95\% \CL limits have been determined by including different combination of uncertainties: statistical plus all systematic uncertainties, statistical plus systematic uncertainties in the fit bias, statistical plus systematic uncertainties in the efficiency. The comparison shows that the systematic uncertainties pertaining to the selection efficiency and the fit bias have similar impact.

The results in terms of the expected 95\% \CL upper limit on the $m_{\Ph}^{\text{mod+}}$ MSSM scenario (with the higgsino mass parameter $\mu = 200$), including the 68 and 95\% \CL bands, are shown in Fig.~\ref{fig:MSSM_hMSSM_limits} (\cmsLeft), in the $m_{\PA}$--$\tan\beta$ plane. The results are obtained including the statistical and all systematic uncertainties. The 95\% \CL upper limit is computed up to $m_{\PA} = 600\GeV$, where the excluded $\tan\beta$ value exceeds 50. For higher values of $\tan\beta$ the MSSM predictions are no longer reliable. These results extend the excluded $\tan\beta$ range obtained at 7 and 8\TeV \cite{bib:cms-mssm-2mu} and also extend the range of the tested $m_{\PA}$ values from 300 to 600\GeV.
The data are also interpreted in terms of the hMSSM model. The corresponding 95\% \CL upper limit on $\tan\beta$ as a function of $m_{\PA}$ are shown in Fig.~\ref{fig:MSSM_hMSSM_limits} (\cmsRight). The observed limits are very similar in the two scenarios, since, in the $m_{\PA}$--$\tan\beta$ range covered by this analysis the $m_{\Ph}^{\text{mod+}}$ predictions for the \Ph boson mass are consistent with the SM Higgs boson mass, and the cross sections of the \PH and \PA bosons are similar between the two models.

The results of the $\Pgt^{+}\Pgt^{-}$ analysis \cite{bib:cms-mssm-tautau} exclude a much larger $m_{\PA}$--$\tan\beta$ region, reaching the value of $\tan\beta = 60$ at $m_{\PA} = 1.5\TeV$. For values of $m_{\PA}$ up to 400\GeV the $\Pgmp\Pgmm$ results exclude a larger $m_{\PA}$--$\tan\beta$ region compared to the results of the $\cPqb\cPaqb$ analysis \cite{bib:cms-mssm-bb}, which is instead slightly more sensitive at higher $m_{\PA}$ reaching the value of $\tan\beta = 60$ at about $m_{\PA} = 700\GeV$.

\begin{figure}
\centering
\includegraphics[width=0.49\textwidth]{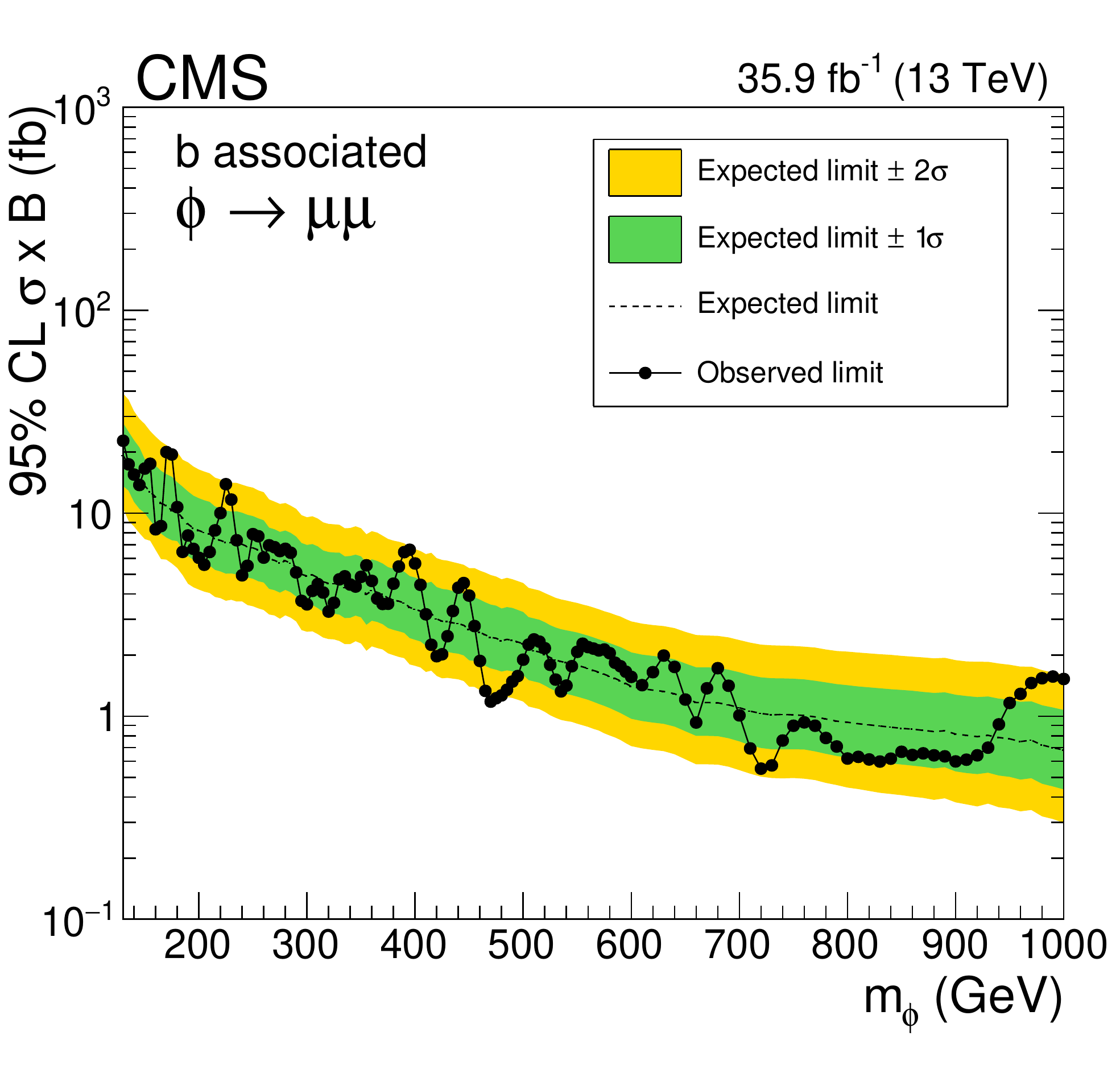}
\includegraphics[width=0.49\textwidth]{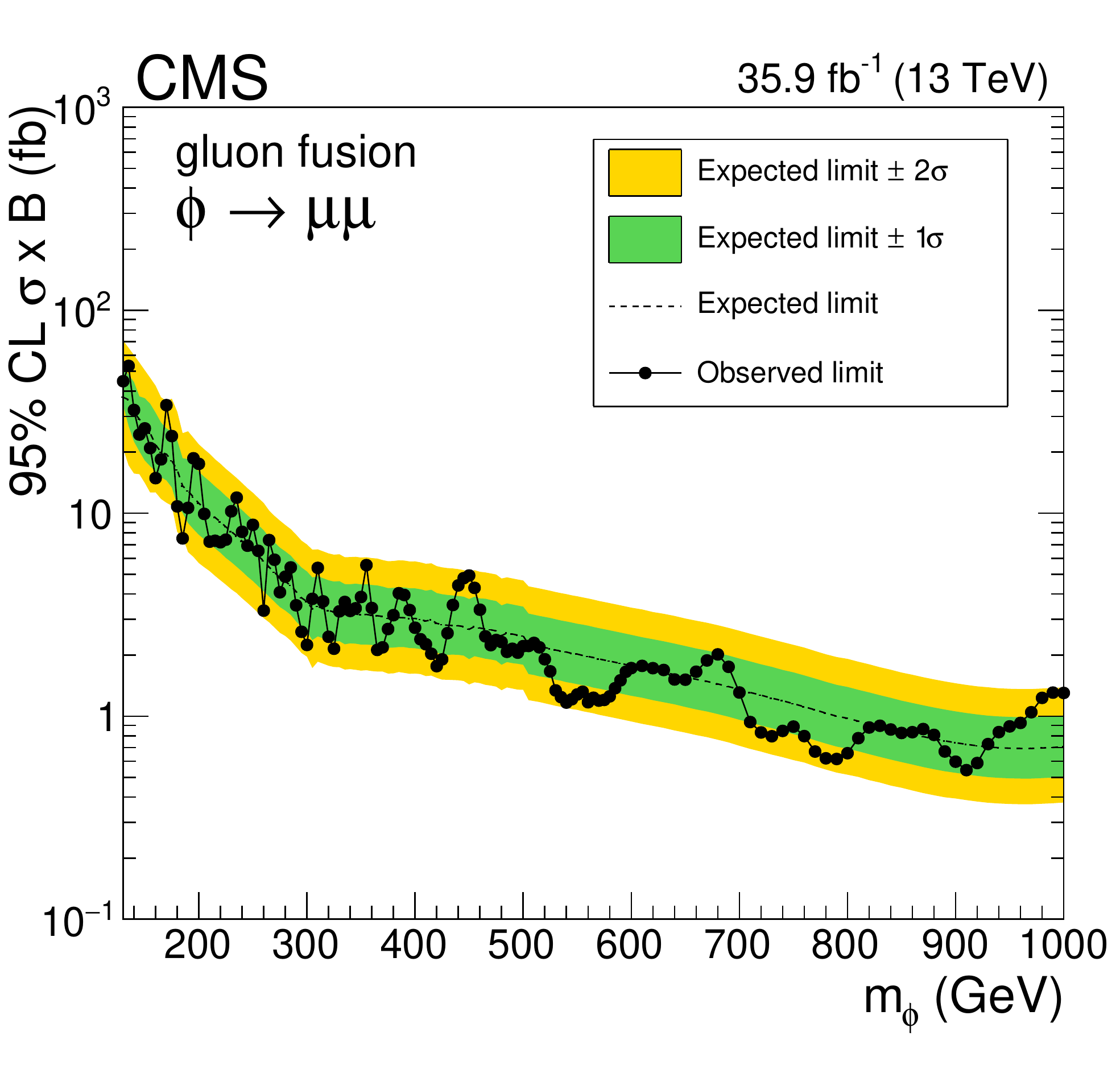}
\caption{The 95\% \CL expected, including the 68 and 95\% \CL bands, and observed model-independent upper limits on the production cross section times branching fraction of a generic $\phi$ boson decaying to a dimuon pair, in the case of \cPqb-associated (\cmsLeft) and gluon fusion (\cmsRight) production. The results are obtained using a signal template with an intrinsic narrow width.}
\label{fig:model_indip}
\end{figure}

\begin{figure}
\centering
\includegraphics[width=0.49\textwidth]{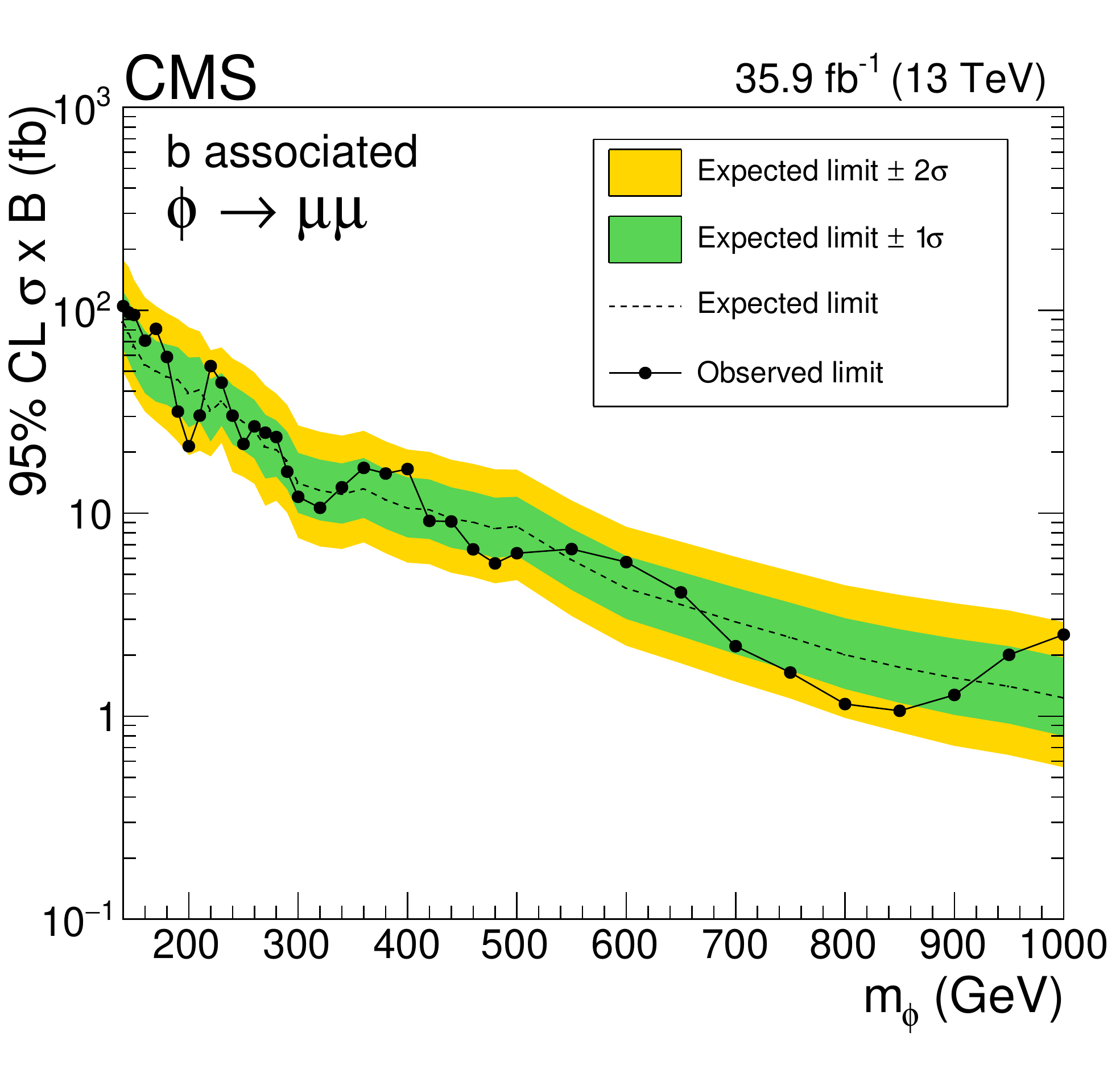}
\includegraphics[width=0.49\textwidth]{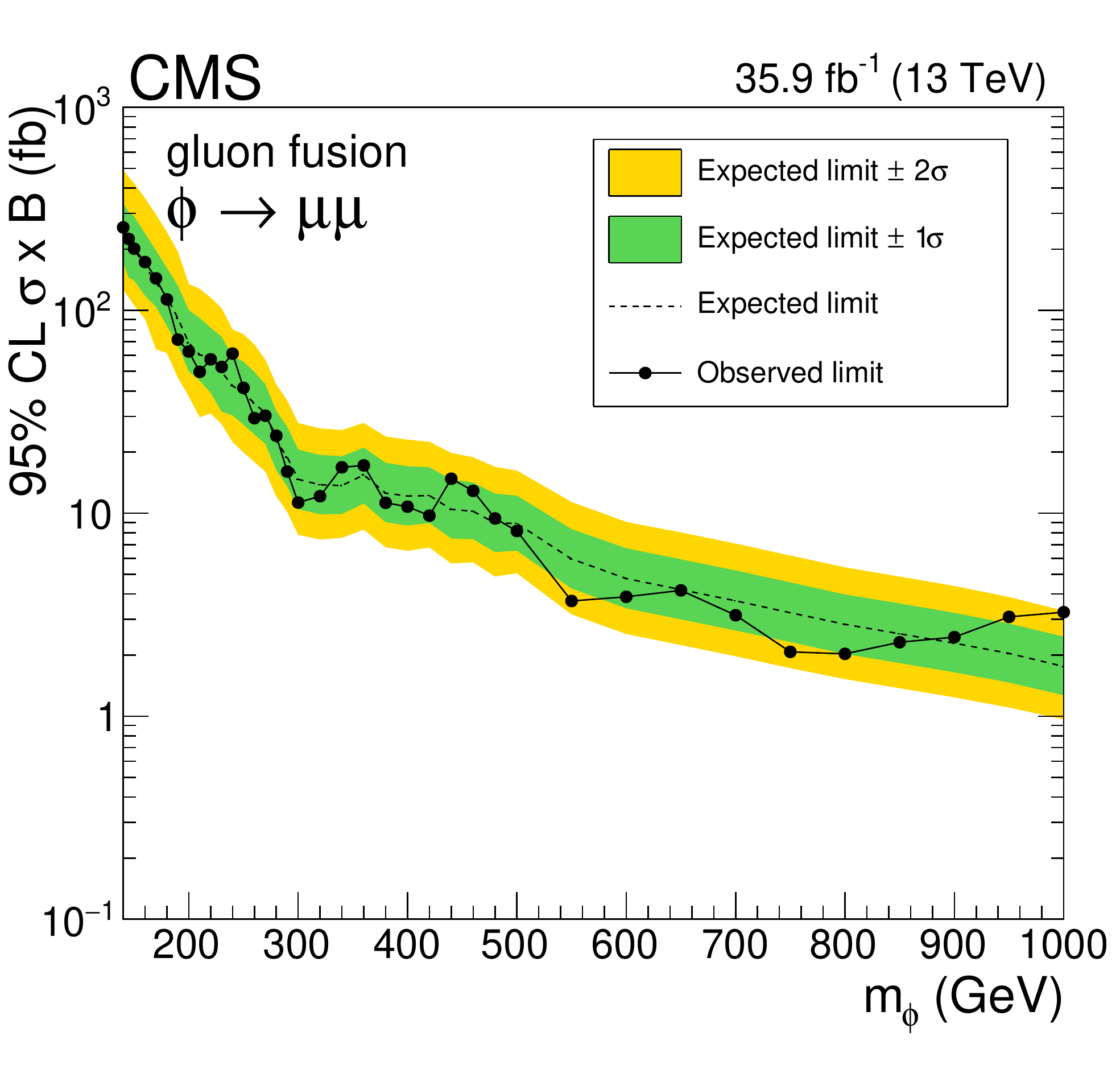}
\caption{The 95\% \CL expected, including the 68 and 95\% \CL bands, and observed model-independent upper limits on the production cross section times branching fraction of a generic $\phi$ boson decaying to a dimuon pair, in the case of \cPqb-associated (\cmsLeft) and gluon fusion (\cmsRight) production. The results are obtained using a signal template with an intrinsic width equal to the 10\% of the nominal mass.}
\label{fig:model_indip_Gamma10percent}
\end{figure}

Limits on the production cross section times decay branching fraction $\sigma\mathcal{B}({\phi}\to\PGmp\PGmm)$ for a single neutral scalar boson $\phi$ have also been determined. In the model-independent interpretation the $\phi$ boson is searched for as a single resonance with mass $m_{\phi}$ assuming a narrow width or a width equal to 10\% of $m_{\phi}$. In the first case the intrinsic width of the signal is smaller than the invariant mass resolution, while in the second case the width is larger even for mass values near 1000\GeV (lower sensitivity of the analysis). The simulated signal of the \PA boson in the $\tan\beta=5$ case (smallest intrinsic width, dominated by the detector resolution) is used as a template to compute the detection efficiency of a generic $\phi$ boson decaying to a muon pair. The $\phi$ boson is assumed to be produced entirely either via the \cPqb-associated or the gluon fusion process, and the analysis is performed separately for the two production mechanisms. Figure~\ref{fig:model_indip} shows the 95\% \CL upper limits on the cross section times the decay branching fraction to $\Pgmp\Pgmm$ as a function of the $\phi$ mass for a narrow resonance. These limits are more stringent by a factor of 2 to 3 than those recently obtained by ATLAS in a similar search \cite{ref:atlas2019}. The corresponding upper limits assuming a signal template with a width equal to 10\% of its mass value are shown in Fig.~\ref{fig:model_indip_Gamma10percent}. In the case of large signal widths, the upper limits as a function of $m_{\phi}$ start from 140\GeV. This is done to have the signal peak $\pm 3\Gamma$ within the fit range. Moreover, as one may expect, the limits are less stringent than for the narrow-width approximation, and it is no longer possible to distinguish the fine structure of the 95\% \CL limits as a function of the mass, as observed for the narrow-width case.  

\section{Summary}
\label{sec:Summary}
A search for neutral minimal supersymmetric standard model (MSSM) Higgs bosons decaying to $\PGmp\PGmm$ was performed using 13\TeV data collected in proton-proton collisions by the CMS experiment at the LHC. No excess of events was found above the expected background due to standard model (SM) processes. The 95\% confidence level upper limit for the production of beyond SM neutral Higgs bosons is determined in the framework of the $m_{\Ph}^{\text{mod+}}$ and the phenomenological scenarios of the MSSM. For the ratio of the vacuum expectation values of the neutral components of the two Higgs doublets, $\tan\beta$, its excluded values range from $\approx$10 to $\approx$60 for a mass of the pseudoscalar \PA boson ($m_{\PA}$) from 130 to 600\GeV. The larger collected luminosity and the higher center-of-mass energy exclude a larger $m_{\PA}$--$\tan\beta$ region, compared to what was obtained at 7 and 8\TeV in a similar analysis. Model-independent exclusion limits on the production cross section times branching fraction of a generic narrow-width neutral boson decaying to two muons have been determined assuming the neutral boson to be produced entirely either via \cPqb-associated or gluon fusion mechanisms. The limits are determined in the mass range from 130 to 1000\GeV, separately for the two production mechanisms. Similarly, exclusion limits are also obtained assuming a signal width equal to 10\% of its mass value.

\begin{acknowledgments}
We congratulate our colleagues in the CERN accelerator departments for the excellent performance of the LHC and thank the technical and administrative staffs at CERN and at other CMS institutes for their contributions to the success of the CMS effort. In addition, we gratefully acknowledge the computing centers and personnel of the Worldwide LHC Computing Grid for delivering so effectively the computing infrastructure essential to our analyses. Finally, we acknowledge the enduring support for the construction and operation of the LHC and the CMS detector provided by the following funding agencies: BMWFW and FWF (Austria); FNRS and FWO (Belgium); CNPq, CAPES, FAPERJ, and FAPESP (Brazil); MES (Bulgaria); CERN; CAS, MoST, and NSFC (China); COLCIENCIAS (Colombia); MSES and CSF (Croatia); RPF (Cyprus); SENESCYT (Ecuador); MoER, ERC IUT, and ERDF (Estonia); Academy of Finland, MEC, and HIP (Finland); CEA and CNRS/IN2P3 (France); BMBF, DFG, and HGF (Germany); GSRT (Greece); NKFIA (Hungary); DAE and DST (India); IPM (Iran); SFI (Ireland); INFN (Italy); MSIP and NRF (Republic of Korea); LAS (Lithuania); MOE and UM (Malaysia); BUAP, CINVESTAV, CONACYT, LNS, SEP, and UASLP-FAI (Mexico); MBIE (New Zealand); PAEC (Pakistan); MSHE and NSC (Poland); FCT (Portugal); JINR (Dubna); MON, RosAtom, RAS and RFBR (Russia); MESTD (Serbia); SEIDI, CPAN, PCTI and FEDER (Spain); Swiss Funding Agencies (Switzerland); MST (Taipei); ThEPCenter, IPST, STAR, and NSTDA (Thailand); TUBITAK and TAEK (Turkey); NASU and SFFR (Ukraine); STFC (United Kingdom); DOE and NSF (USA).

\hyphenation{Rachada-pisek} Individuals have received support from the Marie-Curie program and the European Research Council and Horizon 2020 Grant, contract No. 675440 (European Union); the Leventis Foundation; the A. P. Sloan Foundation; the Alexander von Humboldt Foundation; the Belgian Federal Science Policy Office; the Fonds pour la Formation \`a la Recherche dans l'Industrie et dans l'Agriculture (FRIA-Belgium); the Agentschap voor Innovatie door Wetenschap en Technologie (IWT-Belgium); the F.R.S.-FNRS and FWO (Belgium) under the ``Excellence of Science - EOS" - be.h project n. 30820817; the Ministry of Education, Youth and Sports (MEYS) of the Czech Republic; the Lend\"ulet ("Momentum") Program and the J\'anos Bolyai Research Scholarship of the Hungarian Academy of Sciences, the New National Excellence Program \'UNKP, the NKFIA research grants 123842, 123959, 124845, 124850 and 125105 (Hungary); the Council of Science and Industrial Research, India; the HOMING PLUS program of the Foundation for Polish Science, cofinanced from European Union, Regional Development Fund, the Mobility Plus program of the Ministry of Science and Higher Education, the National Science Center (Poland), contracts Harmonia 2014/14/M/ST2/00428, Opus 2014/13/B/ST2/02543, 2014/15/B/ST2/03998, and 2015/19/B/ST2/02861, Sonata-bis 2012/07/E/ST2/01406; the National Priorities Research Program by Qatar National Research Fund; the Programa Estatal de Fomento de la Investigaci{\'o}n Cient{\'i}fica y T{\'e}cnica de Excelencia Mar\'{\i}a de Maeztu, grant MDM-2015-0509 and the Programa Severo Ochoa del Principado de Asturias; the Thalis and Aristeia programs cofinanced by EU-ESF and the Greek NSRF; the Rachadapisek Sompot Fund for Postdoctoral Fellowship, Chulalongkorn University and the Chulalongkorn Academic into Its 2nd Century Project Advancement Project (Thailand); the Welch Foundation, contract C-1845; and the Weston Havens Foundation (USA).
\end{acknowledgments}

\bibliography{auto_generated}

\cleardoublepage \appendix\section{The CMS Collaboration \label{app:collab}}\begin{sloppypar}\hyphenpenalty=5000\widowpenalty=500\clubpenalty=5000\vskip\cmsinstskip
\textbf{Yerevan Physics Institute, Yerevan, Armenia}\\*[0pt]
A.M.~Sirunyan$^{\textrm{\dag}}$, A.~Tumasyan
\vskip\cmsinstskip
\textbf{Institut für Hochenergiephysik, Wien, Austria}\\*[0pt]
W.~Adam, F.~Ambrogi, T.~Bergauer, J.~Brandstetter, M.~Dragicevic, J.~Erö, A.~Escalante~Del~Valle, M.~Flechl, R.~Frühwirth\cmsAuthorMark{1}, M.~Jeitler\cmsAuthorMark{1}, N.~Krammer, I.~Krätschmer, D.~Liko, T.~Madlener, I.~Mikulec, N.~Rad, J.~Schieck\cmsAuthorMark{1}, R.~Schöfbeck, M.~Spanring, D.~Spitzbart, W.~Waltenberger, J.~Wittmann, C.-E.~Wulz\cmsAuthorMark{1}, M.~Zarucki
% ...
\vskip\cmsinstskip
\dag: Deceased\\
1:  Also at Vienna University of Technology, Vienna, Austria\\
% ...
\end{sloppypar}
\end{document}